\documentclass[prd,twocolumn,groupedaddress,nofootinbib]{revtex4}
\usepackage{graphicx}
\usepackage{latexsym,bm,amsmath,amsfonts,amssymb}
\usepackage{epsfig}

\begin{document}

\title{Neutrino Dark Energy and Baryon Asymmetry from Higgs Sector}

\author{Pei-Hong Gu$^{1}_{}$}

\author{Hong-Jian He$^{2}_{}$}

\author{Utpal Sarkar$^{3}_{}$}

\affiliation{$^{1}_{}$The Abdus Salam International Centre for
Theoretical Physics, Strada Costiera 11, 34014 Trieste, Italy\\
$^{2}_{}$Center for High Energy Physics, Tsinghua University, Beijing 100084, China\\
$^{3}_{}$Physical Research Laboratory, Ahmedabad 380009, India}

\begin{abstract}

We propose a new model to explain the neutrino masses, the dark
energy and the baryon asymmetry altogether. In this model, neutrinos
naturally acquire small Majorana masses via type-II seesaw
mechanism, while the pseudo-Nambu-Goldstone bosons associated with
the neutrino mass-generation mechanism provide attractive candidates
for dark energy. The baryon asymmetry of the universe is produced
from the Higgs triplets decay with CP-violation.

\end{abstract}

\maketitle

\section{Introduction}
The atmospheric, solar and laboratory neutrino oscillation
experiments \cite{pdg2006} have confirmed that neutrinos have tiny
but nonzero masses, of the order of $10^{-2}\,\textrm{eV}$. This
phenomenon is elegantly explained by the seesaw mechanism
\cite{minkowski1977}, in which neutrinos acquire small Majorana
\cite{minkowski1977} or Dirac \cite{mp2002} masses naturally.
Furthermore, the observed matter-antimatter asymmetry in the
universe can be generated through leptogenesis \cite{fy1986} in the
neutrino seesaw scenario.

On the other hand, various cosmological observations \cite{pdg2006}
provide strong evidence that the expansion of our universe is
accelerating due to the existence of dark energy. One possible
explanation for the dark energy has its origin in a dynamical scalar
field, such as the quintessence \cite{wetterich1988} with an
extremely flat potential. It was shown \cite{weiss1987} that the
pseudo-Nambu-Goldstone boson (pNGB) provides an attractive
realization of the quintessence field.

A striking coincidence between the scales of neutrino masses and
dark energy ($\sim(10^{-3}\,\textrm{eV})^4$), inspires us to
consider them in a unified scenario, i.e., the neutrino dark energy
model which generically predicts neutrino-mass variation
\cite{gwz2003}. There have been lots of recent activities studying
the neutrino dark energy models. A possible connection between the
neutrinos and the pNGB dark energy was explored in the type-I seesaw
scenario\,\cite{bhos2005,hmps2006}.

In this paper, we propose a new neutrino dark energy model to
simultaneously explain the generation of neutrino masses and the
origin of dark energy from the Higgs sector. In particular, the
pNGBs associated with neutrino mass-generation provide the
consistent candidates of dark energy while the small neutrino masses
depending on the dark energy field are realized through the type-II
seesaw. Furthermore, the CP-violation and out-of-equilibrium decays
of the Higgs triplets produce the baryon asymmetry in the universe.

\section{The model}
We extend the $SU(2)_{L}^{}\times U(1)_{Y}^{}$ standard model (SM)
with triplet and singlet Higgs scalars,
\begin{eqnarray}
\psi_{Li}^{} &=& \left \lgroup \begin{array}{c}
\nu_{Li}^{}\\
l_{Li}^{}\end{array} \right
\rgroup\,\,(\textbf{2},-\frac{1}{2})\,,\quad \,\,H = \left \lgroup
\begin{array}{c}
H^{0}_{}\\
H^{-}_{}\end{array} \right \rgroup\,\,(\textbf{2},-\frac{1}{2})\,, \nonumber\\
\xi_{ij}^{} &\equiv &\xi_{ji}^{}= \left \lgroup
\begin{array}{cc}
\frac{1}{\sqrt{2}}\delta^{+}_{ij} &\delta^{++}_{ij}\\[1mm]
\delta^{0}_{ij} &-\frac{1}{\sqrt{2}}\delta^{+}_{ij}\end{array}
\right \rgroup\,\,(\textbf{3},1)\,,\nonumber\\
\chi_{ij}^{}&\equiv&\chi_{ji}^{}\,\,(\textbf{1},0)\,,\quad
(\textrm{for}~ i,j=1,2,3),
\end{eqnarray}
are left-handed lepton doublets, Higgs doublet, Higgs triplets and
Higgs singlets, respectively. The full Lagrangian is supposed to be
invariant under a global $U(1)^{6}_{}$ symmetry, generated by the
independent phase transformations of each Higgs triplet among the
six $\xi_{ij}^{}$ fields. Transformations of the Higgs singlets
$\chi_{ij}^{}$ under this $U(1)^{6}_{}$ are determined by requiring
the invariance of the following scalar interactions,
\begin{eqnarray}
\label{yukawa2} \chi_{ij}^{} H^{T}_{} i \tau_{2}^{} \xi_{ij}^{} H +
\textrm{h.c.}\,,
\end{eqnarray}
where we have defined the SM Higgs doublet $H$ as a singlet under
$U(1)^{6}_{}$. The Higgs triplets have the Yukawa couplings to the
left-handed lepton doublets,
\begin{eqnarray}
\label{yukawa1}
\overline{\psi_{Li}^{\,c}}i\tau_{2}^{}\xi_{ij}^{}\psi_{Lj}^{}+\textrm{h.c.}
\,,
\end{eqnarray}
which explicitly break the $U(1)^{6}_{}$ down to its subgroup
$U(1)^{3}_{}$. So we have only three massless Nambu-Goldstone bosons
(NGBs) after the six $\chi_{ij}^{}$'s acquire their vacuum
expectation values (VEVs).

We write down the relevant Lagrangian for the Higgs and
lepton-Yukawa interactions,
\begin{eqnarray}
\label{lagrangian1} \mathcal{L} \supset
&-&\sum_{ij}^{}\left(\overline{\mu}^{2}_{ij}+\sum_{kl}^{}
\lambda^{}_{ij,kl}\chi^{\dagger}_{kl}\chi^{}_{kl}\right)\textrm{Tr}
\left(\xi^{\dagger}_{ij}\xi^{}_{ij}\right)
\nonumber\\
&-&\sum_{ij\neq
kl}^{}\lambda^{'}_{ij,kl}\chi^{\dagger}_{ij}\chi^{}_{kl}
\textrm{Tr}\left(\xi^{\dagger}_{ij}\xi^{}_{kl}\right)
\nonumber\\
&-&\sum_{ij}^{}\left(\frac{1}{2}y^{}_{ij}
\overline{\psi_{Li}^{\,c}}i\tau_{2}^{}\xi_{ij}^{}\psi_{Lj}^{}
-h^{}_{ij}\chi_{ij}^{}H_{}^{T}i\tau_{2}^{}\xi_{ij}^{}H\right.
\nonumber\\
&&\hspace*{10mm}  +\left.\textrm{h.c.}\,\right)\,,
\end{eqnarray}
where $\lambda^{(')}_{ij,kl}\equiv \lambda^{(')}_{ji,kl}\equiv
\lambda^{(')}_{ij,lk}$\,, $y^{}_{ij}\equiv y^{}_{ji}$ and
$h^{}_{ij}\equiv h^{}_{ji}$ are dimensionless while $\mu^{}_{ij}$
has mass-dimension equal one. After the Higgs singlets get their
VEVs, $\langle\chi_{ij}^{}\rangle \equiv
\frac{1}{\sqrt{2}}f_{ij}^{}$\,, we can write
\begin{eqnarray}
\label{chivev} \chi_{ij}^{} =
\frac{1}{\sqrt{2}}\left(f_{ij}^{}+\sigma_{ij}^{}\right)\exp\left(i\varphi_{ij}^{}/f_{ij}^{}\right)\,
\end{eqnarray}
with
$\sigma_{ij}^{}\equiv\sigma_{ji}\,,\,\varphi_{ij}^{}\equiv\varphi_{ji}^{}\,
(i,j=1,2,3)$ being the neutral bosons and the NGBs, respectively.
Among these six NGBs, three of them will acquire nonzero masses via
the Coleman-Weinberg potential (due to the small explicit breaking
of global symmetries, $U(1)^6\to U(1)^3$) and thus become pNGBs. The
other three remain massless as the result of spontaneous breaking of
the subgroup $U(1)^3$.

For convenience, we redefine the Higgs triplets as
\begin{eqnarray}
\label{xinew}
\exp\left(i\varphi_{ij}^{}/f_{ij}^{}\right)\xi_{ij}^{}&\rightarrow&\xi_{ij}^{}\,.
\end{eqnarray}
The mass matrix $M$ for the physical triplet scalar fields is now
given by
\begin{eqnarray}
\label{masstriplet} \hspace*{-7.5mm} M^{}_{ij,kl}
&\!\!\equiv\!\!&\!\!
\left[\!\left(\!\overline{\mu}_{ij}^{2}\!+\!\frac{1}{2}\!\sum_{kl}^{}\!
\lambda_{ij,kl}^{}f_{kl}^{2}\!\right)\!\! \delta_{ij,kl}^{}
+\frac{1}{2}
\lambda^{'}_{ij,kl}f_{ij}^{}f_{kl}^{}\!\right]^{\!\frac{1}{2}}_{}\!\!\!.
\end{eqnarray}
The VEVs of $\chi_{ij}$'s generate the effective trilinear
interactions between the Higgs triplets and doublet,
\begin{eqnarray}
\label{cubic}
\frac{1}{\sqrt{2}}h_{ij}^{}f_{ij}^{}H_{}^{T}i\tau_{2}^{}\xi_{ij}^{}H+\textrm{h.c.}
\equiv \mu_{ij}^{}H_{}^{T}i\tau_{2}^{}\xi_{ij}^{}H+\textrm{h.c.}\,,
\end{eqnarray}
where the cubic couplings $\mu_{ij}^{}$ will be set as real after
proper phase rotations. From Eqs.\,(\ref{xinew}),
(\ref{masstriplet}) and (\ref{cubic}), we derive the Lagrangian
(\ref{lagrangian1}) as below,
\begin{eqnarray}
\label{lagrangian2}\mathcal{L}\supset
&-&\sum_{ij,kl}^{}M^{2}_{ij,kl}\,\textrm{Tr}\left(\xi_{ij}^{\dagger}\xi_{kl}^{}\right)\nonumber\\
&-&\sum_{ij}^{}\left[\frac{1}{2}y^{}_{ij}\exp\left(-i\varphi_{ij}^{}/f_{ij}^{}\right)\overline{\psi_{Li}^{\,c}}i\tau_{2}^{}\xi_{ij}^{}\psi_{Lj}^{}\right.\nonumber\\
&&\hspace*{8mm}-\left.\mu_{ij}^{}H_{}^{T}i\tau_{2}^{}\xi_{ij}^{}H+\textrm{h.c.}\right]\,.
\end{eqnarray}
We still have the freedom to redefine the phases of the three lepton
doublets, which can remove three of the fields $\varphi_{ij}$ from
the lepton-Higgs Yukawa interactions. Without loss of generality, we
choose the rephasing,
\begin{eqnarray}
\label{derivative}
\exp\left[-i\varphi_{ii}^{}/(2f_{ii}^{})\right]\psi_{Li}^{}
&\rightarrow&\psi_{Li}^{}
\end{eqnarray}
which transforms the Lagrangian (\ref{lagrangian2}) into a new form,
 \begin{eqnarray}
 \label{lagrangian3}
\mathcal{L}
\supset&-&\sum_{ij,kl}^{}M^{2}_{ij,kl}\textrm{Tr}\left(\xi_{ij}^{\dagger}\xi_{kl}^{}\right)
\nonumber\\
&-&\left[\frac{1}{2}\right.
\sum_{i}^{}y^{}_{ii}\overline{\psi_{Li}^{\,c}}i\tau_{2}^{}\xi_{ii}^{}\psi_{Li}^{}
\nonumber\\
&&\hspace*{2mm}+\frac{1}{2}\sum_{i\neq
j}^{}y_{ij}^{}\exp(i\phi_{ij}^{}/f)\overline{\psi_{Li}^{\,c}}i\tau_{2}^{}\xi_{ij}^{}\psi_{Lj}^{}
\nonumber\\
&& \hspace*{2mm}
-\sum_{ij}^{}\left.\mu_{ij}^{}H_{}^{T}i\tau_{2}^{}\xi_{ij}^{}H+\textrm{h.c.}\right]
\end{eqnarray}
where
\vspace*{-3mm}
\begin{eqnarray}\label{combination}
\frac{\phi_{ij}^{}}{f} &=& -\frac{\varphi_{ij}^{}}{\,f_{ij}^{}}
+\frac{1}{2}\frac{\varphi_{ii}^{}}{f_{ii}^{}}
+\frac{1}{2}\frac{\varphi_{jj}^{}}{f_{jj}^{}}\,,
\end{eqnarray}
and $f$ is of the order of $f_{ij}^{}$. It is not possible to remove
$\phi_{ij}\,(i\neq j)$ from Eq.\,(\ref{lagrangian3}) by any further
transformations and hence these $\phi_{ij}$'s will become the pNGBs
with tiny masses and can naturally serve as the candidates of dark
energy. Note that from (\ref{xinew}), (\ref{derivative}) and the
subsequent phase rotations on the right-handed charged leptons, the
three massless NGBs, $\varphi_{ii}^{}$, will only have the
derivative interactions to other fields, but they are highly
suppressed by $1/f_{ii}^{}$ and thus escape from experimental
constraints at low energy scales.

\section{Neutrino mass and mixing}
The electroweak symmetry breaking takes
place with the VEV of the Higgs doublet, which induces small VEVs to
the triplets,
\begin{eqnarray}
\langle H\rangle\equiv\frac{1}{\sqrt{2}}\left\lgroup
\begin{array}{c}
v \\
0\end{array} \right\rgroup\!,  && \langle\xi_{ij}^{}\rangle \equiv
\frac{1}{\sqrt{2}}\left\lgroup
\begin{array}{cc}
0 &0\\
u^{}_{ij} &0\end{array} \right\rgroup\!,
\end{eqnarray}
with $i,j=1,2,3$, where the VEVs of the Higgs triplets $\xi_{ij}^{}$
are deduced as
\begin{eqnarray}
\label{vev} u_{ij}^{}&\simeq&
\frac{v^{2}_{}}{\sqrt{2}\,}\sum_{kl}^{}\mu_{kl}^{}\left(M^{-2}_{}\right)_{kl,ij}^{}\,.
\end{eqnarray}
Inspecting Eqs.\,(\ref{masstriplet}) and (\ref{cubic}), it is
natural to take the masses $M_{ij,kl}^{}$ around the same order as
the scalar cubic couplings $\mu_{ij}^{}$ since they are both
controlled by the singlet VEVs ($f_{ij}^{}$).  Hence, the triplet
VEVs in (\ref{vev}) are seesaw-suppressed by the ratio of the
electroweak scale $v$ over the heavy mass $M$, i.e., $u_{ij}^{} =
\mathcal{O}\left(v^{2}_{}/M\right)\ll v$.

These small VEVs will then generate the Majorana masses for the
neutrinos,
\begin{eqnarray}
\label{massneutrino}
\mathcal{L}_{m}&=&-\frac{1}{2}\sum_{ij}^{}
\left(m_{\nu}^{}\right)_{ij}^{}\overline{\nu_{Li}^{\,c}}\nu_{Lj}^{}+\textrm{h.c.}\,,
\end{eqnarray}
\vspace*{-3mm}
where
\begin{eqnarray}
\label{mass1}
\left(m_{\nu}^{}\right)_{ij}^{} = \left\{ \begin{array}{ll} m_{ij}^{}\,,&\quad (\textrm{for}~i=j),\\
m_{ij}^{}\exp\left(i\phi_{ij}^{}/f\right)\,,&\quad (\textrm{for}~i\neq j)\,,\\
\end{array} \right.
\end{eqnarray}
with $\, m_{ij}^{} \equiv
\frac{1}{\sqrt{2}\,}y_{ij}^{}u_{ij}^{}$\,.\, With the $3\times 3$
symmetric mass-matrix (\ref{mass1}), we can readily realize the
neutrino mass-spectrum and mixings, consistent with the neutrino
oscillation experiments. Moreover, the interactions of the neutrinos
with the pNGBs will induce a small long-range force, which can have
direct consequences in cosmology \cite{bbmt2005,rs2006} and neutrino
oscillation experiments \cite{knw2004}.

\section{Origin of dark energy}
\begin{figure}
\vspace{4.0cm} \epsfig{file=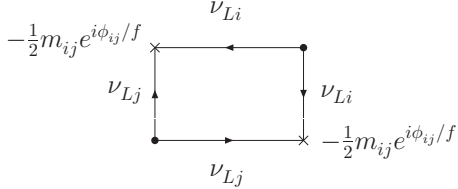, bbllx=6.6cm, bblly=6.0cm,
bburx=16.6cm, bbury=16cm, width=7cm, height=7cm, angle=0, clip=0}
\vspace{-7.5cm} \caption{\label{potential} The one-loop diagram
contributing to the Coleman-Weinberg potential of the pNGBs. }
\end{figure}

Three of the NGBs, $\phi_{ij}^{}\,(i\neq j)$
as defined in (\ref{combination}), will acquire small masses due to
the Yukawa couplings between the left-handed lepton doublets and the
Higgs triplets, and thus become the pNGBs. As shown in Fig.
\ref{potential}, the leading loop diagram will contribute a
Coleman-Weinberg effective potential for $\phi_{ij}^{}$. Similar to
\cite{bhos2005}, we explicitly calculate the potential,
\begin{eqnarray}
V(\phi_{12}^{},\phi_{23}^{},\phi_{31}^{})=-\frac{1}{32\pi^{2}_{}}\sum_{k=1}^{3}m_{k}^{4}\ln
\frac{m_{k}^{2}}{\Lambda^{2}_{}}\,,
\end{eqnarray}
where $m_{k}^{}$ as a function of $\phi_{ij}^{}$ is the \textit{k}th
eigenvalue of the neutrino mass matrix $m_{\nu}^{}$, and $\Lambda$
is the ultraviolet cutoff. A typical term in $V$ that contributes to
the potential of a pNGB field $Q$ has the form,
\begin{eqnarray}
V(Q)\simeq V_{0}^{}\cos (Q/f)
\end{eqnarray}
with $V_{0}^{}=\mathcal{O}(m_{\nu}^{4})$. It is well known that with
$f$ of the order of Planck mass $M_{\textrm{Pl}}^{}$, the pNGB $Q$
will acquire a mass of order
$\mathcal{O}(m_{\nu}^{2}/M_{\textrm{Pl}}^{})$ and thus provides a
consistent candidate for the quintessence dark energy.

Finally, we also note that after the electroweak symmetry breaking,
the explicit breaking of the $U(1)^3$ symmetry is only generated by
the dimension-3 soft mass term in Eq.\,(15). So, in this model there
is no dimension-4 term which explicitly breaks $U(1)^3$, and thus
there are no higher order correction that could contribute a
divergent term to spoil the renormalizability and the original
symmetry of the theory. This point can be checked by explicit
calculations as well. For instance, at the two-loop order, we have
\begin{eqnarray}
V_{2}^{}&\approx&
\left(\frac{1}{16\pi^{2}_{}}\right)^{2}_{}\,\Lambda^{2}_{}\,\textrm{Tr}
\left(m_{\nu}^{}m_{\nu}^{\dagger}\right)\,\left[\textrm{Tr}
\left(y_{}^{}y_{}^{\dagger}\right)\right.\nonumber\\
&&+\left.\left(\frac{1}{2}
+\frac{1}{4\cos^{2}_{}\theta_{W}^{}}\right)g^{2}_{}\right]\,.
\nonumber
\end{eqnarray}
Since $\textrm{Tr}\left( m_{\nu}^{}m_{\nu}^{\dagger}\right)$ is
independent on the pNGBs as well as $\textrm{Tr}
\left(y_{}^{}y_{}^{\dagger}\right)$, we see that the two-loops have
no contribution to the effective potential for the pNGBs. Similarly,
there is no contribution from other higher loops.

\section{Baryon asymmetry}

\begin{figure*}
\vspace{4.0cm} \epsfig{file=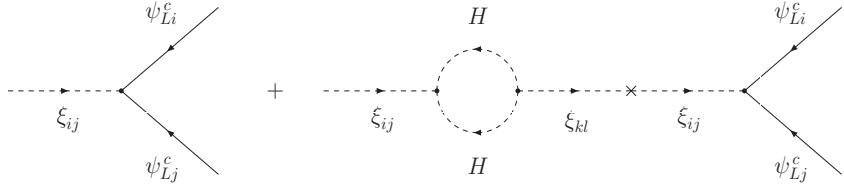, bbllx=6cm, bblly=6.0cm,
bburx=16cm, bbury=16cm, width=6.1cm, height=6.3cm, angle=0, clip=0}
\vspace{-6.5cm} \caption{\label{decay} The Higgs triplets decay to
the leptons at one-loop order. }
\end{figure*}

The decays of Higgs triplets,
\begin{eqnarray}
\xi_{ij}^{} &\rightarrow& \left\{ \begin{array}{ll}
\psi_{Li}^{\,c}\,\psi_{Lj}^{\,c}\,, &\,\,\,\,\,(\,L\,=\, -\,2\,)\,,  \\
H^{\ast}_{}\,H^{\ast}_{}\,, &\,\,\,\,\,(\,L\,=\,0\,)\,,
\end{array} \right.
\end{eqnarray}
can break the lepton number. As shown in Fig.\,\ref{decay}, the
mass-mixings in (\ref{masstriplet}) among different Higgs triplets
contribute the tree-level and one-loop diagrams that interfere to
generate CP asymmetry in these decays. The decays of the triplet
Higgs will then produce enough lepton asymmetry before the
electroweak phase transition, which can successfully explain the
observed matter-antimatter asymmetry in the universe through the
sphaleron processes\,\cite{krs1985} which convert the lepton
asymmetry into the existing baryon asymmetry.

To calculate the CP-asymmetry, we define the Higgs triplets in their
mass-eigenbasis,
\begin{eqnarray}
\widehat{\xi}_{a}^{}&\equiv&\sum_{ij}^{}\,U_{a,ij}^{}\xi_{ij}^{}
\end{eqnarray}
with the diagonalized mass-eigenvalues,
\begin{eqnarray}
\widehat{M}_{a}^{}&\equiv&\sum_{ij,kl}^{}U_{a,ij}^{}M_{ij,kl}^{}U^{}_{a,kl}\,.
\end{eqnarray}
Thus we can rewrite the Lagrangian (\ref{lagrangian3}) as
\begin{eqnarray} \label{lagrangian4} \mathcal{L} &\supset&
-\sum_{a}^{}\widehat{M}^{2}_{a}\textrm{Tr}
\left(\widehat{\xi}_{a}^{\,\dagger}\widehat{\xi}_{a}^{}\right)
-\left(\frac{1}{2}\right.\sum_{a,ij}^{}
\widehat{y}^{a}_{ij}\overline{\psi_{Li}^{\,c}}i\tau_{2}^{}
\widehat{\xi}_{a}^{}\psi_{Lj}^{}
\nonumber\\
&& -\sum_{a}^{}\left.\widehat{\mu}_{a}^{}H_{}^{T}i\tau_{2}^{}
\widehat{\xi}_{a}^{}H+\textrm{h.c.}\right)\,
\end{eqnarray}
with the definition \vspace*{-2mm}
\begin{eqnarray}
\widehat{\mu}_{a}^{}&\equiv&\sum_{ij}^{}U_{a,ij}^{}\mu_{ij}^{}
\end{eqnarray}
\vspace*{-2.5mm} and
\begin{eqnarray}
\label{yukawa} \widehat{y}^{a}_{ij} \equiv
\left\{ \begin{array}{ll} U_{a,ij}^{}y_{ij}^{}\,,&(\textrm{for}~ i=j)\,,\\
U_{a,ij}^{}y_{ij}^{}\exp\left(i\phi_{ij}^{}/f\right)\,,&(\textrm{for}~
i\neq j)\,.
\\
\end{array} \right.
\end{eqnarray}
Here $U$ is an orthogonal rotation matrix. The proper CP-asymmetry
parameter is then described by \cite{ms1998}
\begin{eqnarray}
\label{cpv} \varepsilon_{a}^{}&\equiv&
 2\!\times\!\frac{\sum_{ij}^{}\left[\Gamma\!\left(\widehat{\xi}^{\ast}_{a}\!\rightarrow\!
\psi_{Li}^{}\psi_{Lj}^{}\!\right)-\Gamma\!\left(\widehat{\xi}_{a}^{}\!\rightarrow\!
\psi_{Li}^{\,c}\psi_{Lj}^{\,c}\!\right)\right]}{\Gamma_{a}^{}}\nonumber\\
&=&\frac{1}{\pi}\sum_{b\neq a}^{}\frac{\textrm{Im}\left\{
\textrm{Tr}\left[\left(\widehat{y}^{b}_{}
\right)^{\dagger}_{}\widehat{y}^{a}_{}\right]\right\}
\widehat{\mu}_{b}^{}\widehat{\mu}_{a}^{}}
{\textrm{Tr}\left[\left(\widehat{y}^{a}_{}\right)^{\dagger}_{}
\widehat{y}^{a}_{}\right]\hat{M}_{a}^{2}+4\,\widehat{\mu}_{a}^{2}}
\frac{\widehat{M}_{a}^{2}}
{\widehat{M}_{a}^{2}-\widehat{M}_{b}^{2}}\,
\end{eqnarray}
with
\\[-8mm]
\begin{eqnarray}
\Gamma_{a}^{}=
\frac{1}{8\pi}\left\{\frac{1}{4}\textrm{Tr}\left[\left(\widehat{y}^{a}_{}\right)^{\dagger}_{}
\widehat{y}^{a}_{}\right]
+\frac{\widehat{\mu}_{a}^{2}}{\widehat{M}_{a}^{2}}\right\}
\widehat{M}_{a}^{}\,
\end{eqnarray}
being the total decay width of $\widehat{\xi}_{a}^{}$ or
${\widehat\xi}_{a}^{\,\ast}$.

For illustration, we use $\widehat{\xi}_{a}^{}$ to denote the
lightest Higgs triplet and hence its contribution is expected to
dominate the final baryon asymmetry, which is given by the
approximate relation \cite{kt1980},
\begin{eqnarray}
\label{asymmetry} Y_{B}^{}\equiv\frac{n_{B}^{}}{s}\simeq
-\frac{28}{79}\times\left\{
\begin{array}{ll}
\displaystyle \frac{\varepsilon_{a}^{}}{g_{\ast}^{}}\,,
&~(\textrm{for}~K \ll 1)\,,
\\[4mm]
\displaystyle \frac{0.3\,\varepsilon_{a}^{}}{g_{\ast}^{}K\left(\ln
K\right)^{0.6}_{}}\,, &~(\textrm{for}~K \gg 1)\,,
\end{array} \right.
\end{eqnarray}
with the factor $28/79$ being the value of $B/(B-L)$ and the
parameter $K$ defined by
\begin{eqnarray}
\label{parameter} K &\equiv& \left.
\frac{\Gamma_{a}^{}}{2H(T)}\right|_{T=\widehat{M}_{a}^{}}\,
\end{eqnarray}
as a measurement of the departure from equilibrium. Here
$H(T)=(8\pi^{3}_{}g_{\ast}^{}/90)^{\frac{1}{2}}T^{2}_{}/M_{\textrm{Pl}}^{}$
is the Hubble constant with the Planck mass
$M_{\textrm{Pl}}^{}\simeq 1.2\times 10^{19}_{}\,\textrm{GeV}$ and
the relativistic degrees of freedom $g_{\ast}^{}\simeq 106.75$. For
simplicity, we further consider
\\[-5mm]
\begin{eqnarray}
\label{assumption}
r_{ba}^{}\equiv\frac{\widehat{M}_{b}^{}}{\widehat{M}_{a}^{}}=\frac{\widehat{\mu}_{b}^{}}{\widehat{\mu}_{a}^{}}\gg
1\quad\textrm{and}\quad \widehat{y}^{b}_{}\equiv
\widehat{y}^{a}_{}e^{-i\delta_{ba}^{}}_{}\,,
\end{eqnarray}
where $\delta_{ba}$ is the relative phase between
$\widehat{y}^{b}_{}$ and $\widehat{y}^{a}_{}$. We thus neglect the
contribution from the heavier triplets to the neutrino masses and
conveniently express $K$ as
\\[-5mm]
\begin{eqnarray}
\label{kexpress}
K=\left[\frac{(4\pi)^{5}_{}g_{\ast}^{}}{45}\right]^{-\frac{1}{2}}_{}\left(B_{\psi}^{}B_{H}^{}\right)^{-\frac{1}{2}}_{}\frac{M_{\textrm{Pl}}^{}\overline{m}}{v^{2}_{}}\,.
\end{eqnarray}
Here the quadratic mean of the neutrino masses $(\overline{m})$ is
defined by
\\[-5mm]
\begin{eqnarray}
\label{quadraticmass}
\overline{m}^{2}_{}&\equiv&\sum_{k=1}^{3}m_{k}^{2}\equiv
\textrm{Tr}\left(m_{\nu}^{\dagger}m_{\nu}^{}\right)\simeq
\frac{1}{4}\textrm{Tr}\left[\left(\widehat{y}^{a}_{}\right)^{\dagger}_{}\widehat{y}^{a}_{}\right]
\frac{\widehat{\mu}_{a}^{2}v^{4}_{}}{\widehat{M}_{a}^{4}}\nonumber\\
&=&(8\pi)^{2}_{}B_{\psi}^{}B_{H}^{}\Gamma_{a}^{2}\frac{v^{4}_{}}{\widehat{M}_{a}^{4}}
\end{eqnarray}
and $(B_{\psi}\,,\,B_{H})$ are the branching ratios of the
tree-level decays of $\widehat{\xi}_{a}^{}$ into the lepton and
Higgs doublets, which always hold the relationship,
\begin{eqnarray}
B_{\psi}^{}+B_{H}^{}\equiv 1\,,\quad \Longrightarrow\quad
B_{\psi}^{}B_{H}^{}\leqslant \frac{1}{4}\,.
\end{eqnarray}
Then, we compute the CP-asymmetry (\ref{cpv}) as
\begin{eqnarray}
\label{cpv2} \varepsilon_{a}^{} &=&-\frac{1}{\pi}\frac{\textrm{Tr}
\left[\left(\widehat{y}^{a}_{}\right)^{\dagger}_{}\widehat{y}^{a}_{}\right]
\widehat{\mu}_{a}^{2}}{\textrm{Tr}
\left[\left(\widehat{y}^{a}_{}\right)^{\dagger}_{}\widehat{y}^{a}_{}\right]
\widehat{M}_{a}^{2}+4\widehat{\mu}_{a}^{2}}
\sum_{b\neq a}^{}\frac{r_{ba}^{}\sin\delta_{ba}^{}}{r_{ba}^{2}-1}\,\nonumber\\
&\simeq&-\frac{1}{\pi}\left(B_{\psi}^{}B_{H}^{}\right)^{1/2}_{}
\frac{\widehat{M}_{a}^{}\overline{m}}{v^{2}_{}}\frac{\sin\delta_{ba}^{}}{r_{ba}^{}}\,.
\end{eqnarray}
Inputting\, $v\simeq 246\,\textrm{GeV}$,\, $B_{\psi}^{}B_{H}^{}=
1/4$,\, $\widehat{M}_{a}^{}= 4\times 10^{12}$ GeV,\, $\overline{m}=
0.1\,\textrm{eV}$,\, $\sin\delta_{ba}^{} = 0.1$\, and\, $r_{ba}^{} =
10$,\, we derive the sample predictions: $\,\varepsilon_{a}^{}
\simeq -1.1\times 10^{-5}_{}\,$\, and $\,K\simeq 46\,$. Finally, we
deduce, \,$n_{B}^{}/s\simeq 10^{-10}$,\, consistent with the
cosmological observations.

\section{Conclusion and discussion}

In this paper, we propose a new model to unify the neutrino dark
energy and baryon asymmetry by extending only the SM Higgs sector
with triplet and singlet Higgs scalars. The Higgs triplets naturally
acquire tiny VEVs and give small Majorana masses for the neutrinos.
The model contains three pNGBs associated with the neutrino
mass-generation, which provide consistent candidates for the
quintessence field. The matter-antimatter asymmetry in the universe
is produced by the out-of-equilibrium decays of the Higgs triplets
with CP-violating couplings. We can readily accommodate the dark
matter as well in our construction by adding a darkon field
\cite{sz1985} or an inert Higgs doublet \cite{bhr2006}.

In our model, the neutrino masses are functions of the dark energy
field. Being a dynamical component, the dark energy will evolute
with time and/or in space. Accordingly, the neutrino masses will
vary instead of being constants. The prediction of the neutrino-mass
variation could be verified in the present and future experiments,
such as the observations on the cosmic microwave background and the
large scale structures \cite{bbmt2005}, the measurement of the
extremely high-energy cosmic neutrinos \cite{rs2006}, and the
analysis of the neutrino oscillation data \cite{knw2004}.

\vspace*{2mm} \textbf{Acknowledgments}: P.H.G. would like to thank
Jun-Bao Wu, Wander G. Ney and Alexei Yu. Smirnov for helpful
discussions.

\end{document}